\documentstyle[amssymb,preprint,aps]{revtex}

\begin{document}
\title{{\bf Green's function for a n-dimensional closed, static universe and with a
spherical boundary}}
\author{Mustafa \"{O}zcan}
\address{Department of Physics, Trakya University 22030 Aysekadin/Edirne-TURKEY}
\date{June 25, 2001}
\maketitle

\begin{abstract}
We construct the Hadamard Green's function by using the eigenfunction, which
are obtained by solving the wave equation for the massless conformal scalar
field on the $S^{n-1}$ of a n-dimensional closed, static universe. We also
consider the half space case with both the Dirichlet and the Neumann
boundary conditions. Solving of eigenfunction and eigenvalues of the
corresponding field equation is interesting since the Casimir energy could
be calculated analytically by various methods.
\end{abstract}

\section{\ INTRODUCTION}

In 1948, Casimir \cite{casi} showed that one consequence of the zero-point
quantized field is an attractive force between two uncharged, perfectly
conducting parallel plates. The principal message of this force is that
changes in infinite vacuum energy of the quantized field can be finite and
observable \cite{spaarnay,lam}.

The Casimir energy leads to negative energy between the plates, depends on
the distance between plates and independent plate area, as well as similar
calculation were made for sphere \cite{boyer}, for which the Casimir energy
was found to be positive which means that the force on the shell is
repulsive and depends only on the radius of the shell. The type of Casimir
energy values depend on manifolds of different topology and geometry \cite
{mostenenko}. The boundary and/or curvature conditions play the important
role of the quantized field.

The Casimir effect has also been calculated in curved background spacetime.
In 1975, Ford \cite{ford} showed that the vacuum energy of a massless
conformal scalar field in an Einstein universe by the mode sum method. Ford
obtained the renormalized vacuum energy density in this case as

\begin{equation}
\rho =\frac{\text{%
h\hskip-.2em\llap{\protect\rule[1.1ex]{.325em}{.1ex}}\hskip.2em%
c}}{480\pi ^{2}R_{0}^{4}},  \label{wave}
\end{equation}

where, $R_{0}$ is the radius of the universe and the pressure is given by $P=%
\frac{1}{3}\rho .$

Thus the energy momentum tensor is in the same form as that for the
classical radition. In Ford's work on the Casimir effect in an Einstein
universe, he showed that the energy is associated with the closed spatial
topology. Ford used the mode sum method and employed an exponential cutoff
in his calculations. Later, Dowker and Critchley \cite{dowker} verified
Ford's result by using the covariant point-splitting method. Their
calculation was not only covariant but also cutoff independent. Even though
Dowker and Critchley used the global topology of the universe to construct
the Green's function, their result is also expected to be true locally;
i.e., the local renormalized energy density and pressure should be given by

\begin{equation}
\left( \left\langle T_{0}^{0}\right\rangle =\right) \text{ \ }\rho =\frac{%
\hbar c}{480\pi ^{2}R_{0}^{4}}\text{ , \ }\left( -\left\langle
T_{1}^{1}\right\rangle =-\left\langle T_{2}^{2}\right\rangle =-\left\langle
T_{3}^{3}\right\rangle =\right) \text{ P=}\frac{1}{3}\rho \text{ .}
\end{equation}

Three often used regularization techniques of the quantum vacuum energy in a
curved background spacetimes are the mode sum method \cite{ford}, the zeta
function regularization technique \cite{elizalde} and the point-splitting
method. One of them, the must powerful method available is the
point-splitting regularization technique \cite{birrel book,fulling book}.
This method works with point separated functions. In the Hamiltonian, the
scalar field $\Phi \left( x\right) $ behaves like a field operator due to
the canonical quantization scheme. However, the products of field operators
and their covariant derivatives are at the same point which causes the
vacuum expectation value of the energy momentum tensor to the diverge. To
eliminate these divergent quantities, one of the field operators $\Phi
\left( x\right) $ in each product in the Hamiltonian replaced by $\Phi
\left( x^{\prime }\right) $, where $x^{\prime }$ is some point near $x.$
Thus the separeted function $\langle 0\left| \Phi \left( x\right) \Phi
\left( x^{\prime }\right) \right| 0\rangle $ transforms like the product two
functions, one at $x$, the other at $x^{\prime }$. The two point function $%
\left\langle 0\left| \Phi \left( x\right) \Phi \left( x^{\prime }\right)
\right| 0\right\rangle $ is called the positive Wightmann function and is
denoted by

\begin{equation}
G^{\left( +\right) }\left( x,x^{\prime }\right) =\left\langle 0\left| \Phi
\left( x\right) \Phi \left( x^{\prime }\right) \right| 0\right\rangle .
\end{equation}

The vacuum expectation value of the anticommutator of the field is called
the Hadamard Green's function, which is defined by \cite{birrel book}

$\ \ \ \ \ \ \ \ \ \ \ \ \ \ \ \ \ \ \ \ \ \ \ \ \ \ \ \ \ \ \ \ \ \ $%
\begin{equation}
G^{\left( 1\right) }\left( x,x^{\prime }\right) =\left\langle 0\left|
\left\{ \Phi \left( x\right) \Phi \left( x^{\prime }\right) +\Phi \left(
x^{\prime }\right) \Phi \left( x\right) \right\} \right| 0\right\rangle ,
\end{equation}

where $G^{\left( 1\right) }\left( x,x^{\prime }\right) $ and $G^{\left(
+\right) }\left( x,x^{\prime }\right) $ are related by

\begin{equation}
G^{\left( 1\right) }\left( x,x^{\prime }\right) =2%
\mathop{\rm Re}%
G^{\left( +\right) }\left( x,x^{\prime }\right) .
\end{equation}

In this work, we use the above physical motivation for the massless
conformal scalar field in a n-dimensional closed, static universe. Firstly,
we will construct the Hadamard Green's function for the case of a
n-dimensional static, closed universe, using the separated function idea.
Secondly, we construct the Green's function directly by using the wave
function satisfying the boundary conditions. The half space case (as denoted
by Kennedy and Unwin \cite{kennedy}) is very important in the sense that
despite being in curved spacetime, and with a spherical boundary. Our
results confirm the Green's function for the full n-dimensional closed,
static universe as an image sum and obtained the necessary Green's function
for the n-dimensional half space case by locating an image charge in dual
region.

We have organized this paper as follows. In section II. we calculate the
mode function for the massless conformal scalar field in a n- dimensional
closed, static universe. In section III. we constructed the Hadamard Green's
function by using the eigenfunction, which are obtained by solving wave
equation for the massless conformal scalar field on $S^{n-1}$ ($n\geq 4,$ $n$
is the dimension of the spacetime). In section IV. we study the Green's
function for the massless conformal scalar field with a spherical boundary
at $\chi _{0}=\frac{\pi }{2}$ and with the interior geometry represented by
the closed, static n-dimensional universe (this case is also called the half
universe). We consider both the Dirichlet and the Neumann boundary
conditions.

\section{\protect\bigskip THE MODE FUNCTION}

We shall use hyperspherical polar coordinates $\left( \chi ,\theta ,\phi
\right) =\left( \chi ,\theta _{1},\theta _{2},\theta _{3},.....,\theta _{\mu
},\phi \right) $ ; $\mu =1,2,3,.....,n-3.$ Where $n$\bigskip\ is the
spacetime dimension. The line element in the time orthogonal form

\begin{equation}
ds^{2}=dt^{2}-R_{0}^{2}dl^{2},
\end{equation}

\bigskip where $R_{0}$ is the radius of the universe and is a constant. In a
static-closed model the part $dl^{2}$ follows by imagining that the space
sections of fixed world time $t$ are embedded in a n-dimensional space with
coordinates $x_{1},x_{2},x_{3},...,x_{n-1},w$ and line element $%
dl^{2}=dx_{1}^{2}+dx_{2}^{2}+.......+dx_{n-1}^{2}+dw^{2}.$ Our $n-1$
dimensional space is the surface of the sphere $%
x_{1}^{2}+x_{2}^{2}+.......+x_{n-1}^{2}+w^{2}=R_{0}^{2}$ at fixed $R_{0}.$
Spherical coordinates for this space are

\begin{eqnarray}
x_{1} &=&R_{0}\sin \chi \sin \theta _{1}\sin \theta _{2}.........\sin \theta
_{n-3}\cos \phi , \\
x_{2} &=&R_{0}\sin \chi \sin \theta _{1}\sin \theta _{2}.........\sin \theta
_{n-3}\sin \phi ,  \nonumber \\
x_{3} &=&R_{0}\sin \chi \sin \theta _{1}\sin \theta _{2}.........\cos \theta
_{n-3},  \nonumber \\
&&\vdots  \nonumber \\
x_{n-1} &=&R_{0}\sin \chi \cos \theta _{1},  \nonumber \\
w &=&R_{0}\cos \chi .  \nonumber
\end{eqnarray}

Where $\chi \in \left[ 0,\pi \right] $, $\ \theta _{\mu }\in \left[ 0,\pi %
\right] $ $\ \mu =1,2,3,...,n-3$, and $\phi \in \left[ 0,2\pi \right] $.

Thus, the general metric for a static, closed n-dimensional spacetimes has
the form

\begin{eqnarray}
ds^{2} &=&dt^{2}-R_{0}^{2}[d\chi ^{2}+\sin ^{2}\chi d\theta _{1}^{2}+\sin
^{2}\chi \sin ^{2}\theta _{1}d\theta _{2}^{2}+..........  \nonumber \\
&&+\sin ^{2}\chi \sin ^{2}\theta _{1}\sin ^{2}\theta
_{2}................\sin ^{2}\theta _{n-4}d\theta _{n-3}^{2}  \nonumber \\
&&+\sin ^{2}\chi \sin ^{2}\theta _{1}\sin ^{2}\theta _{2}.....\sin
^{2}\theta _{n-4}\sin ^{2}\theta _{n-3}d\phi ^{2}].
\end{eqnarray}

We shall consider conformal massless scalar field $\Phi \left( x\right) $ on
this curved background geometry. It satisfies the field equation

\begin{equation}
\square \Phi \left( x\right) +\xi \left( n\right) R\Phi \left( x\right) =0.
\end{equation}

(We follow the sign convention in \cite{birrel book}) Where $\xi \left(
n\right) =\frac{1}{4}\frac{\left( n-2\right) }{\left( n-1\right) }$ is the
dimensional parameter determining the coupling between $\Phi \left( x\right) 
$ and the Ricci scalar R. The Ricci scalar for the metric Eq.$\left(
8\right) $ is given by

\begin{equation}
R=\frac{1}{R_{0}^{2}}\left( n-2\right) \left( n-1\right) \text{ \ for }n\geq
2.
\end{equation}

We solve Eq. (9) using the method of separation of variables. In general,
for $n\geq 4$ we can write

\begin{eqnarray}
\Phi \left( x\right) &=&c_{0n}\text{ }e^{-i\omega t}\text{ }X\left( \chi
\right) X_{1}\left( \theta _{1}\right) X_{2}\left( \theta _{2}\right)
......X_{n-4}\left( \theta _{n-4}\right) Y_{lm}\left( \theta _{n-3},\phi
\right)  \nonumber \\
\text{ \ \ \ \ \ \ \ with }\omega &=&\frac{N}{R_{0}}\text{ .}
\end{eqnarray}
\ 

Where $Y_{lm}$ is the spherical harmonic and $c_{0n}$ is appropriate
normalization constant, and

\bigskip 
\begin{equation}
\left( N_{2}=\right) \medskip l=0,1,2,3,........,\text{and \ \ \ \ \ \ }%
m=-l,-l+1,...,0,1,......,l-1,l\text{ \ .}
\end{equation}

We now calculate the mode function for $n\geq 4$. The $X\left( \chi \right) $
and $X_{\mu }\left( \theta _{\mu }\right) $ ($\mu =1,2,3,.......,n-4$ )
satisfy the following differential equations

\[
\frac{1}{\sin ^{n-4}\chi }\frac{d}{d\chi }\left( \sin ^{n-2}\chi \frac{d}{%
d\chi }X\left( \chi \right) \right) \text{ \ \ \ \ \ \ \ \ \ \ \ \ \ \ \ \ \
\ \ \ \ \ \ \ \ \ \ \ \ \ \ \ \ }
\]
\ \ \ \ \ \ \ \ \ \ \ \ \ \ \ \ \ \ \ \ \ \ \ \ \ \ 
\begin{equation}
\medskip \text{ \ \ \ \ \ \ \ \ \ \ \ \ \ \ \ \ }+\left[ \left( N^{2}-\frac{%
\left( n-2\right) ^{2}}{4}\right) \sin ^{2}\chi -N_{n-2}\left(
N_{n-2}+n-3\right) \right] X\left( \chi \right) =0\text{ \ , and}
\end{equation}

\[
\frac{1}{\sin ^{n-\mu -4}\theta _{\mu }}\frac{d}{d\theta _{\mu }}\left( \sin
^{n-\mu -2}\theta _{\mu }\frac{d}{d\theta _{\mu }}X_{\mu }\left( \theta
_{\mu }\right) \right) \text{ \ \ \ \ \ \ \ \ \ \ \ \ \ \ \ \ \ \ \smallskip
\medskip\ \ \ \ \ \ \ \ \ \ \ \ \ \ \ \ \ \ \ \ \ \ \ \ \ \ \ \ \ } 
\]
\ \ \ \ \ \ \ \ \ \ \ \ \ \ \ \ \ \ \ \ \ \ \ \ \ \ \ \ \ \ \ \ \ \ \ \ \ \
\ \ \ \ \ \ \ \ \ \ \ \ \ \ \ \ \ \ \ \ \ \ \ \ \ \ \ \ \ \ \ \ \ \ \ \ \ \
\ \ \ \ \ \ \ \ \ \ \ \ \ \ \ \ \ \ \ \ \ \ \ \ \ \ \ \ \ \ \ \ \ \ \ \ \ \
\ \ \ \ \ \ \ \ \ \ \ \ \ \ \ \ \ \ \ \ \ \ \ \ \ \ \ \ \ \ \ \ \ \ \ \ \ \
\ \ \ \ \ \ \ \ \ \ \ \ \ \ \ \ \ \ \ \ \ \ \ \ \ \ \ \ \ \ \ \ \ \ \ \ \ \
\ \ \ \ \ \ \ \ \ \ \ \ \ \ \ \ \ \ \ \ \ \ \ \ \ \ \ \ \ \ \ \ \ \ \ \ 

\begin{equation}
\text{ \ \ }\medskip \text{ \ \ \ \ \ \ }+\left[ N_{n-\mu -1}\left( N_{n-\mu
-1}+n-\mu -2\right) \sin ^{2}\theta _{\mu }-N_{n-\mu -2}\left( N_{n-\mu
-2}+n-\mu -3\right) \right] X_{\mu }\left( \theta _{\mu }\right) =0
\end{equation}

$\bigskip $with \ $\mu =1,2,3,4,.........,n-4.$

Where $($ $N_{2}=)l,N_{3,}N_{4,}N_{5},.......N_{n-3},N_{n-2}$ and \ $N=R_{0}$
$\omega $ \ are the separation parameters.

Notice that we still have not calculated the regularity of the solution in
our intervals at $\theta _{\mu }\epsilon \left[ 0,\pi \right] $ \ where $\mu
=1,2,3,......,n-4$; hence $N_{j}\left( j=3,4,5,...,n-2\right) $ and $%
N=R_{0}\omega $ remain continuous parameters while $l$ and $m$ take the
values $\medskip (N_{2}=)l=0,1,2,3,4,5,6,........,$ \ \ \ \ \ \ $%
m=-l,-l+1,...,0,1,......,l-1,l$ \ .

Making the following substitutions for each one in Eqs. (13) and (14) \ 

\[
X\left( \chi \right) \propto \sin ^{N_{n-2}}X\left( \chi \right) C_{1}\left(
\cos \chi \right) \text{ \ , and} 
\]

\begin{eqnarray}
X_{\mu }\left( \theta _{\mu }\right) \text{ } &\propto &\text{ }\sin
^{N_{n-\mu -2}}\theta _{\mu }\text{ }C_{\mu }\left( \cos \theta _{\mu
}\right) \text{ \ \medskip\ } \\
\text{with }\mu &=&1,2,3,4,......,n-4\text{.}  \nonumber
\end{eqnarray}

We obtain\ \ \ \ \ \ \ \ \ \ \ \ \ \ \ \ \ \ \ \ 

\[
\left( 1-x^{2}\right) \frac{d^{2}}{dx^{2}}C\left( x\right) -x\left(
2N_{n-2}+n-1\right) \frac{d}{dx}C\left( x\right) \medskip \ \ 
\]

\begin{equation}
\text{ \ \ \ \ \ \ \ \ \ \ \ }+\left[ \left( N^{2}-\frac{\left( n-2\right)
^{2}}{4}\right) -N_{n-2}\left( N_{n-2}+n-2\right) \right] C\left( x\right) =0%
\text{ ,}
\end{equation}

\bigskip where $x=\cos \chi $.

\[
\text{ \ \ \ \ \ }\left( 1-x_{\mu }^{2}\right) \frac{d^{2}}{dx_{\mu }^{2}}%
C_{\mu }\left( x_{\mu }\right) -x_{\mu }\left( 2N_{n-\mu -2}+n-\mu -1\right) 
\frac{d}{dx_{\mu }}C_{\mu }\left( x_{\mu }\right) \text{ \ \ \ \ \ \ \ \ \ \
\ \ \ \ \ \ \ \ \ \ \ \ \ \ \ \ \ \ \ \ \ }\medskip 
\]

\begin{equation}
+\left[ N_{n-\mu -1}\left( N_{n-\mu -1}+n-\mu -2\right) -N_{n-\mu -2}\left(
N_{n-\mu -2}+n-\mu -2\right) \right] C_{\mu }\left( x_{\mu }\right)
=0\medskip ,
\end{equation}

\bigskip where $x_{\mu }=\cos \theta _{\mu }$ , \ $\mu =1,2,3,4,.......,n-4$
.\bigskip

\bigskip

The differential equation Eq. (17) can be solved by the method of Frobenius.
All solution sets are also divergent at the end points of our interval. To
obtain regular solutions in the entire interval $x_{\mu }\in \left[ -1,1%
\right] $ $\left( \mu =1,2,3,4,........,n-4\right) $ we terminate the
infinite series after a finite number of terms by restricting \ $%
N_{3},N_{4},............,N_{n-2}$ to have integer values.

\begin{eqnarray}
N_{n-2} &=&0,1,2,3,4,.............\text{ \ \ \ ,}  \nonumber \\
N_{n-\mu -1} &=&0,1,2,3,4,................,N_{n-\mu }\text{ with \ }\mu
=2,3,4,...,n-4\text{,}  \nonumber \\
\left( N_{2}=\right) l &=&0,1,2,3,4,.........,N_{3}  \nonumber \\
m &=&-l,-l+1,.......,0,1,......,l\text{ \ .}
\end{eqnarray}

The polynomial solution obtained this way for the equation (17) is the
Gegenbauer polynomials

\begin{equation}
X_{\mu }\left( \theta _{\mu }\right) \propto \text{ sin}^{N_{n-\mu
-2}}\theta _{\mu }\text{ C}_{N_{n-\mu -1}-N_{n-\mu -2}}^{N_{n-\mu -2}+\frac{%
n-\mu -2}{2}}\left( \cos \theta _{\mu }\right) \text{ \ with }\mu
=1,2,3,4,...,n-4\text{ .}
\end{equation}

We return to Eq. (16). This differential equation can be solved by the
method of Frobenius. Hence, we express the function $C\left( x\right) $ \ by
a power series as

\begin{equation}
C\left( x\right) =\sum_{k=0}^{\infty }\text{ }a_{k}x^{k+\alpha }\text{ .}
\end{equation}

This leads us to the following recursion relation:

\begin{equation}
\frac{a_{k+2}}{a_{k}}=\frac{\left[ k+N_{n-2}+\frac{\left( n-2\right) ^{2}}{2}%
\right] -N^{2}}{\left( k+1\right) \left( k+2\right) }\text{ .}
\end{equation}

The above recursion relation gives us the following series solution for $%
C\left( x\right) $ :

\begin{equation}
C\left( x\right) =\sum_{k=0}^{\infty }\text{ }a_{2k}\text{ }%
x^{2k}+\sum_{k=0}^{\infty }\text{ }a_{2k+1}\text{ }x^{2k+1}\text{ .}
\end{equation}

Using the Recurrence relation this is in the form

\begin{equation}
C\left( x\right) =a_{0}C_{1}\left( x\right) +a_{1}C_{2}\left( x\right) \text{
,}
\end{equation}

Where $C_{1}\left( x\right) $ and $C_{2}\left( x\right) $ are linearly
independent. Convergence of the series at the end points of our interval can
be checked eaisly by using the Raabe test which says that if $\ 
\mathop{\textstyle\sum}%
\limits_{k=0}^{\infty }$ $u_{k}$ is a series of positive terms and if the $%
\lim\limits_{n\rightarrow \infty }n\left( \frac{u_{n}}{u_{n+1}}-1\right) =A$
the series is divergent for $A<1$ , convergent for $A>1$ and the test fails
for $A=1$ $\cite{arfken}$. At the end points we apply this test to the
series with even powers i.e. $C_{1}\left( x\right) .$ We write

\begin{equation}
\frac{a_{2k}}{a_{2k+1}}=\frac{\left( 2k+1\right) \left( 2k+2\right) }{\left[
2k+N_{n-2}+\frac{\left( n-2\right) }{2}\right] ^{2}-N^{2}}\text{ ,}
\end{equation}

\bigskip

one obtains $A=\frac{3}{2}-N_{n-2}-\frac{\left( n-2\right) }{2}$ \ $\left( 
\text{with N}_{n-2}\geq 0\text{ and }n\geq 4\right) $ $A<1$ . Hence the
infinite series $C_{1}\left( x\right) $ \ diverges at the end points. One
may similarly check that $C_{2}\left( x\right) $ is also divergent at the
end points of the interval. To obtain a regular solution in the entire
interval $x\in \left[ -1,1\right] $ \ we terminate the infinite series after
a finite number of terms by restricting $N$ \ to have integer values given as

\begin{equation}
N\left( k,N_{n-2}\right) =k+N_{n-2}+\frac{\left( n-2\right) }{2}\text{ ,}
\end{equation}

where \ $k=0,1,2,3,4,.......$ \ and $N_{n-2}=0,1,2,3,..........$ \ \ with \ $%
\omega =\frac{N_{n-2}}{R_{0}}$\ . Redefining a new index $k^{\prime }$ as $%
k+N_{n-2\text{ }}$ \ we write \ $N=k^{\prime }+\frac{\left( n-2\right) }{2}$
\ with \ $\omega _{k}=\frac{N}{R_{o}}$ \ and \ $k^{\prime }=0,1,2,3,....,$ \
and \ $N_{n-2}=0,1,2,3,....,k^{\prime }$ . Dropping primes one obtains the
eigenfrequencies as

\begin{equation}
\omega _{k}=\frac{1}{R_{0}}\left[ k+\frac{\left( n-2\right) }{2}\right] 
\text{ \ \ with \ }k=0,1,2,3,4,...\text{.... \ \ .}
\end{equation}

Values that $N\left( k,N_{n-2}\right) $ could take are presented in table I..

For fixed $k$, one obtains the degeneracy of each eigenfrequencies

\begin{equation}
g_{k}=\sum\limits_{N_{n-2}=0}^{k}\sum_{N_{n-3}=0}^{N_{n-2}}...........%
\sum_{N_{5}=0}^{N_{6}}\sum_{N_{4}=0}^{N_{5}}\sum_{N_{3}=0}^{N_{4}}%
\sum_{l=0}^{N_{3}}\left( 2l+1\right) \text{ \ .}
\end{equation}

The polynomial solutions obtained this way are the well known Gegenbauer
polynomials, hence we can write $C_{1}\left( x\right) $ in terms of the
Gegenbauer polynomials as $C_{1}\left( x\right) =C_{k-N_{n-2}}^{N_{n-2}+%
\frac{\left( n-2\right) }{2}}\left( \cos \chi \right) $.

Thus, the mode solution for $n\geq 4$ becomes

\begin{eqnarray}
\Phi _{\lambda }\left( x\right) &=&c_{0n}\text{ e}^{-i\omega _{k}t}\text{ sin%
}^{N_{n-2}}\chi \text{ }C_{k-N_{n-2}}^{N_{n-2}+\frac{\left( n-2\right) }{2}%
}\left( \cos \chi \right) \medskip  \nonumber \\
&&%
\mathop{\displaystyle\prod}%
\limits_{\mu =1}^{n-4}\left[ \sin ^{N_{n-\mu -2}\text{ }}\theta _{\mu }\text{
\ }C_{N_{n-\mu -1}-N_{n-\mu -2}}^{N_{n-\mu -2}+\frac{n-\mu -2}{2}}\text{ }%
\left( \cos \theta _{\mu }\right) \right] \text{ }Y_{lm}\left( \theta
_{n-3,}\phi \right) \text{ ,}
\end{eqnarray}

\bigskip with

\begin{eqnarray}
\omega _{k} &=&\frac{1}{R_{0}}\left[ k+\frac{\left( n-2\right) }{2}\right] 
\text{ , \ where }k=0,1,2,3,4,......\text{ \ \ \ ,}  \nonumber \\
N_{n-2} &=&0,1,2,3,4,.......,k  \nonumber \\
N_{n-\left( \mu +1\right) } &=&0,1,2,3,4,............,N_{n-\mu }\text{ \
with \ }\mu =2,3,4,...n-4\text{ \ ,}  \nonumber \\
\left( N_{2}=\right) l &=&0,1,2,3,4,..............,N_{3}\text{ \ \ \ \ and} 
\nonumber \\
m &=&-l,-l+1,.....,0,1,2,.....,l\text{ \ \ ,\ }
\end{eqnarray}

and $c_{0n}$ is the normalization constant.

To evaluate the normalization constant we write the scalar product

\begin{eqnarray}
\left( \Phi _{\lambda _{1},}\Phi _{\lambda _{2}}\right) &=&-i\int \left[
\Phi _{\lambda _{1}}\left( x\right) \text{ }\partial _{t}\text{ }\Phi
_{\lambda _{2}}^{\ast }\left( x\right) -\partial _{t}\text{ }\Phi _{\lambda
_{1}}\left( x\right) \text{ }\Phi _{\lambda _{2}}^{\ast }\left( x\right) %
\right] \sqrt{-g}\text{ }d^{n-1}x  \nonumber \\
&=&\delta _{\lambda _{1}\lambda _{2}\text{ }}\text{ .}
\end{eqnarray}

$\Phi _{\lambda }\left( \lambda \text{ stands for }%
k,N_{n-2},N_{n-3},......,N_{4},N_{3},(N_{2}=)l\text{ and }m\text{ values
which are given in Eq. (29)}\right) $ given Eq. (28)\ and, from a complete
orthonormal set with respect to the product Eq. (30). Using Eq. (30) we
obtain $c_{0n}$ as

\begin{eqnarray}
c_{0n} &=&\frac{2^{N_{n-2}+\frac{\left( n-4\right) }{2}}}{R_{0}^{\frac{n-2}{2%
}}}\text{ }\Gamma \left( N_{n-2}+\frac{\left( n-2\right) }{2}\right) \left[ 
\frac{\left( k-N_{n-2}\right) !}{\pi \Gamma \left( k+N_{n-2}+n-2\right) }%
\right] ^{\frac{1}{2}}  \nonumber \\
&&%
\mathop{\displaystyle\prod}%
\limits_{\mu =1}^{n-4}\left\{ 2^{N_{n-\mu -2}+\frac{\left( n-\mu -3\right) }{%
2}}\text{ }\Gamma \left( N_{n-\mu -2}+\frac{\left( n-\mu -2\right) }{2}%
\right) \right.  \nonumber \\
&&\left. \left[ \frac{\left( N_{n-\mu -1}-N_{n-\mu -2}\right) !\left(
N_{n-\mu -1}+\frac{\left( n-\mu -2\right) }{2}\right) }{\pi \Gamma \left(
N_{n-\mu -1}+N_{n-\mu -2}+n-\mu -2\right) }\right] ^{\frac{1}{2}}\right\} 
\text{.}
\end{eqnarray}

The general solution of the field equation can be written as sum over these
modes in the form

\begin{equation}
\Phi \left( x\right) =%
\mathop{\displaystyle\sum}%
\limits_{\lambda }\left( a_{\lambda }\text{ }\Phi _{\lambda }\left( x\right)
+a_{\lambda }^{\dagger }\Phi _{\lambda }^{\ast }\left( x\right) \right) 
\text{ .}
\end{equation}

Where, $a_{\lambda }$ is an annihilation operator and $a_{\lambda }^{\dagger
}$'s a creation operator for quanta in mode $\lambda $, while $\Phi
_{\lambda }\left( x\right) $ is the solution of the wave equation given by
Eq. (28).

\section{THE GREEN'S FUNCTION}

The Green's function we aim to calculate is defined by

\begin{equation}
\left\langle 0\left| \text{ }\Phi \left( x\right) \text{ }\Phi \left(
x^{\prime }\right) \right| 0\right\rangle =%
\mathop{\displaystyle\sum}%
\limits_{\lambda }\text{ }\Phi _{\lambda }\left( x\right) \Phi _{\lambda
}^{\ast }\left( x^{\prime }\right) \text{ ,}
\end{equation}

where $\Phi _{\lambda }\left( x\right) $\ are the mode function.

Using the mode solutions given in equation one obtains

\bigskip 
\begin{eqnarray}
G^{\left( +\right) }\left( x,x^{\prime }\right) &=&\frac{1}{\pi R_{0}^{n-2}}%
\mathop{\displaystyle\sum}%
\limits_{k=0}^{\infty }e^{-i\frac{\Delta t}{R_{0}}\left( k+\frac{\left(
n-2\right) }{2}\right) }%
\mathop{\displaystyle\sum}%
\limits_{N_{n-2}=0}^{k}2^{2N_{n-2}+n-4}\left\{ \Gamma \left( N_{n-2}+\frac{%
\left( n-2\right) }{2}\right) \right\} ^{2}  \nonumber \\
&&\frac{\left( k-N_{n-2}\right) !}{\Gamma \left( N_{n-2}+k+n-2\right) }\text{%
sin}^{N_{n-2}}\chi \text{sin}^{N_{n-2}}\chi ^{\prime }\text{ }%
C_{k-N_{n-2}}^{N_{n-2}+\frac{\left( n-2\right) }{2}}\left( \cos \chi \right)
\medskip \text{ }C_{k-N_{n-2}}^{N_{n-2}+\frac{\left( n-2\right) }{2}}(\cos
\chi ^{\prime })  \nonumber \\
&&\text{\ \ \ \ \ \ }%
\mathop{\displaystyle\prod}%
\limits_{\mu =1}^{n-4}\left\{ 
\mathop{\displaystyle\sum}%
\limits_{N_{n-\mu -2}=0}^{N_{n-\mu -1}}2^{2N_{n-\mu -2}+n-\mu -3}\left\{
\Gamma \left( N_{n-\mu -2}+\frac{\left( n-\mu -2\right) }{2}\right) \right\}
^{2}\text{ }\right.  \nonumber \\
&&\left[ \frac{\left( N_{n-\mu -1}-N_{n-\mu -2}\right) !\left( N_{n-\mu -1}+%
\frac{\left( n-\mu -2\right) }{2}\right) }{\pi \Gamma \left( N_{n-\mu
-1}+N_{n-\mu -2}+n-\mu -2\right) }\right]  \nonumber \\
&&\text{\ \ \ }\left. \text{sin}^{N_{n-\mu -2}}\theta _{\mu }\text{ sin}%
^{N_{n-\mu -2}}\theta _{\mu }^{\prime }\text{ }C_{N_{n-\mu -1}-N_{n-\mu
-2}}^{N_{n-\mu -2}+\frac{\left( n-\mu -2\right) }{2}}\left( \cos \theta
_{\mu }\right) \medskip \text{ }C_{N_{n-\mu -1}-N_{n-\mu -2}}^{N_{n-\mu -2}+%
\frac{\left( n-\mu -2\right) }{2}}\left( \cos \theta _{\mu }^{\prime
}\right) \right\}  \nonumber \\
&&\text{\ \ \ }%
\mathop{\displaystyle\sum}%
\limits_{m=-l}^{l}\text{ }Y_{lm\text{ }}\left( \theta _{n-3},\phi \right) 
\text{ }Y_{lm}^{\ast }\left( \theta _{n-3}^{\prime },\phi ^{\prime }\right) 
\text{ .}
\end{eqnarray}

Using the addition formula for the spherical harmonics \cite{arfken}, which
is given as

\begin{eqnarray}
\frac{\left( 2l+1\right) }{4\pi }P_{l}\left( \cos \text{ }\gamma _{1}\right)
&=&%
\mathop{\displaystyle\sum}%
\limits_{m=-l}^{l}\text{ }Y_{lm\text{ }}\left( \theta _{n-3},\phi \right) 
\text{ }Y_{lm}^{\ast }\left( \theta _{n-3}^{\prime },\phi ^{\prime }\right) 
\text{ ,}  \nonumber \\
\text{where }\cos \text{ }\gamma _{1} &=&\cos \theta _{n-3}\cos \theta
_{n-3}^{\prime }+\sin \theta _{n-3}\sin \theta _{n-3}^{\prime }\cos \left(
\phi -\phi ^{\prime }\right) \text{ .}
\end{eqnarray}

And, using the $n-3$ times addition theorem for the Gegenbauer polynomials 
\cite{tablebook}

\[
C_{\nu }^{p}\left( \cos \vartheta \cos \varphi +\sin \vartheta \sin \varphi
\cos \psi \right) =\frac{\Gamma \left( 2p-1\right) }{\left\{ \Gamma \left(
p\right) \right\} ^{2}}%
\mathop{\displaystyle\sum}%
\limits_{\eta =0}^{\nu }2^{2\eta }\left\{ \Gamma \left( p+\eta \right)
\right\} ^{2}\frac{\left( \nu -\eta \right) !\left( 2\eta +2p-1\right) }{%
\Gamma \left( \nu +\eta +2p\right) } 
\]

\begin{equation}
\text{ \ \ \ \ \ \ \ \ \ \ \ \ \ \ \ \ \ \ \ \ \ \ \ \ \ \ \ \ \ \ \ \ \ \ \
\ \ \ \ \ \ \ \ \ \ \ \ \ \ \ \ sin}^{\eta }\vartheta \text{sin}^{\eta
}\varphi \text{ }C_{\nu -\eta }^{p+\eta }\left( \cos \vartheta \right)
\medskip \text{\ }C_{\nu -\eta }^{p+\eta }\left( \cos \varphi \right) \text{ 
}C_{\eta }^{p-\frac{1}{2}}\left( \cos \psi \right) \text{.}
\end{equation}

We obtain

\begin{equation}
G^{\left( +\right) }\left( x,x^{\prime }\right) =\frac{\alpha _{n}}{4\pi
^{n-2}R_{0}^{n-2}}%
\mathop{\displaystyle\sum}%
\limits_{k=0}^{\infty }\text{ }e^{-i\frac{\Delta t}{R_{0}}\left( k+\frac{%
\left( n-2\right) }{2}\right) }\text{ }C_{k}^{\frac{\left( n-2\right) }{2}%
}\left( \cos \gamma _{n-2}\right) \text{ ,}
\end{equation}

where

\begin{equation}
\alpha _{n}=2^{0}.2^{1}.2^{2}....2^{n-4}\frac{\left\{ \Gamma \left( 1\right)
\right\} ^{2}\left\{ \Gamma \left( \frac{3}{2}\right) \right\} ^{2}\left\{
\Gamma \left( 2\right) \right\} ^{2}.......\left\{ \Gamma \left( \frac{n-2}{2%
}\right) \right\} ^{2}}{\Gamma \left( 1\right) \Gamma \left( 2\right) \Gamma
\left( 3\right) \Gamma \left( 4\right) .................\Gamma \left(
n-3\right) }\text{ ,}
\end{equation}

and

\begin{eqnarray*}
\cos \text{ }\gamma _{n-2} &=&\cos \chi \cos \chi ^{\prime }+\sin \chi \sin
\chi ^{\prime }\cos \gamma _{n-3}\text{ ,} \\
\cos \text{ }\gamma _{n-\mu -2} &=&\cos \theta _{\mu }\cos \theta _{\mu
}^{\prime }+\sin \theta _{\mu }\sin \theta _{\mu }^{\prime }\cos \gamma
_{n-\mu -3} \\
\text{with \ }\mu &=&1,2,3,4,......,n-4\text{ \ , and}
\end{eqnarray*}

\begin{equation}
\text{ \ \ \ \ \ \ \ \ \ \ \ }\cos \text{ }\gamma _{1}=\cos \theta
_{n-3}\cos \theta _{n-3}^{\prime }+\sin \theta _{n-3}\sin \theta
_{n-3}^{\prime }\cos \left( \phi -\phi ^{\prime }\right) \text{ \ .}
\end{equation}

Substituting $\Delta t-i\epsilon $ for with the understanding that we will
let $\epsilon \rightarrow 0$ at the end we get $\left| e^{-i\frac{\Delta t}{%
R_{0}}}\right| <1$ . Now, we could use the generating function for the
Gegenbauer polynomials, one obtains

\begin{equation}
G^{\left( +\right) }\left( x,x^{\prime }\right) =\frac{\alpha _{n}}{4\pi
^{n-2}R_{0}^{n-2}}\frac{1}{\left( \cos \frac{\Delta t}{R_{0}}-\cos \gamma
_{n-2}\right) ^{\frac{\left( n-2\right) }{2}}}\text{ .}
\end{equation}

Where $\gamma _{n-2}$ is the angle between the vectors ${\bf r}$ and ${\bf r}%
^{\prime }$ in the direction $\left( \chi ,\theta _{1},\theta _{2},\theta
_{3},......,\theta _{n-3},\phi \right) $ and $\left( \chi ^{\prime },\theta
_{1}^{\prime },\theta _{2}^{\prime },\theta _{3}^{\prime },......,\theta
_{n-3}^{\prime },\phi ^{\prime }\right) $ . The geodesic distance on $S^{n-1}
$ is denoted by $\Delta s_{n-2}\left( q,q^{\prime }\right) =R_{0}\gamma
_{n-2}$ where $q=q\left( \chi ,\theta _{1},\theta _{2},\theta
_{3},......,\theta _{n-3},\phi \right) $ ; $q^{\prime }$ and $q\in $ $S^{n-1}
$ and $x=x\left( t,q\right) $ . The spacelike separation $\Delta
s_{n-2}\left( q,q^{\prime }\right) $ is given in terms of the coordinates $%
\left( \chi ,\theta _{1},\theta _{2},\theta _{3},......,\theta _{n-3},\phi
\right) $ and $\left( \chi ^{\prime },\theta _{1}^{\prime },\theta
_{2}^{\prime },\theta _{3}^{\prime },......,\theta _{n-3}^{\prime },\phi
^{\prime }\right) $ as

\begin{equation}
\left( \cos \gamma _{n-2}=\right) \text{ \ cos}\frac{\Delta s_{n-2}}{R_{0}}%
=\cos \chi \cos \chi ^{\prime }+\sin \chi \sin \chi ^{\prime }\cos \gamma
_{n-3}\text{ ,}
\end{equation}

where $\cos $ $\gamma _{n-3}$ is given in Eq. (39).

Thus, we have the following the two point function

\begin{equation}
\left\langle 0\left| \Phi \left( x\right) \text{ }\Phi \left( x^{\prime
}\right) \right| 0\right\rangle =\frac{\alpha _{n}}{4\pi ^{n-2}R_{0}^{n-2}}%
\frac{1}{\left( \cos \frac{\Delta t}{R_{0}}-\cos \frac{\Delta s_{n-2}}{R_{0}}%
\right) ^{\frac{\left( n-2\right) }{2}}}\text{ \ .}
\end{equation}

Notice that the Hadamard Green's function $G^{\left( +\right) }\left(
x,x^{\prime }\right) $ is related to $G^{\left( 1\right) }\left( x,x^{\prime
}\right) $ through the relation 
\begin{eqnarray}
G^{\left( 1\right) }\left( x,x^{\prime }\right) &=&2{\bf 
\mathop{\rm Re}%
}\text{ }\left\langle 0\left| \Phi \left( x\right) \text{ }\Phi \left(
x^{\prime }\right) \right| 0\right\rangle \text{ ,}  \nonumber \\
&=&2{\bf 
\mathop{\rm Re}%
}\text{ }G^{\left( +\right) }\left( x,x^{\prime }\right) \text{ .}
\end{eqnarray}

\section{\protect\bigskip THE GREEN'S FUNCTION FOR HALF UNIVERSE}

We will now investigate the massless conformal scalar field with a spherical
boundary located at $\chi _{0}=\frac{\pi }{2}$ \ and with interior geometry
represented by the static closed metric. The metric inside the boundary may
be expressed as

\begin{eqnarray}
ds^{2} &=&dt^{2}-R_{0}^{2}[d\chi ^{2}+\sin ^{2}\chi d\theta _{1}^{2}+\sin
^{2}\chi \sin ^{2}\theta _{1}d\theta _{2}^{2}+..........  \nonumber \\
&&+\sin ^{2}\chi \sin ^{2}\theta _{1}\sin ^{2}\theta
_{2}................\sin ^{2}\theta _{n-4}d\theta _{n-3}^{2}  \nonumber \\
&&+\sin ^{2}\chi \sin ^{2}\theta _{1}\sin ^{2}\theta _{2}.....\sin
^{2}\theta _{n-4}\sin ^{2}\theta _{n-3}d\phi ^{2}]\text{ ,}
\end{eqnarray}

\bigskip

\bigskip where $\chi \in \left[ 0,\frac{\pi }{2}\right] ,$ \ $\theta _{\mu
}\in \left[ 0,\pi \right] ,$ \ $\mu =1,2,3,4,5,....,n-3$ \ and \ $\phi \in %
\left[ 0,2\pi \right] .$

We consider a massless conformal scalar field on this curved background
geometry where the wave equation is given by

\begin{equation}
\square \Phi \left( x\right) +\xi \left( n\right) R\Phi \left( x\right) =0.
\end{equation}

Where $\xi \left( n\right) =\frac{1}{4}\frac{\left( n-2\right) }{\left(
n-1\right) }$ is the conformal coupling constant and $R$ is the Ricci
scalar: $R=\frac{1}{4}\left( n-2\right) \left( n-1\right) $ for $n\geq 4$ ($%
n $ is the dimension of spacetime).

Solution of Eq. (45) could be found easily

\begin{eqnarray}
\Phi _{\lambda }\left( x\right) &=&\overline{c_{0n}}\text{ e}^{-i\omega t}%
\text{ sin}^{N_{n-2}}\chi \text{ }C_{k-N_{n-2}}^{N_{n-2}+\frac{\left(
n-2\right) }{2}}\left( \cos \chi \right) \medskip  \nonumber \\
&&%
\mathop{\displaystyle\prod}%
\limits_{\mu =1}^{n-4}\left[ \sin ^{N_{n-\mu -2}\text{ }}\theta _{\mu }\text{
\ }C_{N_{n-\mu -1}-N_{n-\mu -2}}^{N_{n-\mu -2}+\frac{\left( n-\mu -2\right) 
}{2}}\text{ }\left( \cos \theta _{\mu }\right) \right] \text{ }Y_{lm}\left(
\theta _{n-3,}\phi \right) \text{ ,}
\end{eqnarray}

where $\overline{c_{0n}}$ is the normalization constant, and \ $\omega =%
\frac{N(k)}{R_{0}}$ .

We note that the boundary condition at \ $\chi =\chi _{0}$ has not imposed
on Eq. (46) yet. Hence, $N$ still remains a contiuous parameter, while $%
N_{n-2},N_{n-\left( \mu +1\right) }$ $\left( \mu =2,3,4,...,n-4\right) ,$ $l$
and $m$ take the values.

\begin{eqnarray}
N_{n-2} &=&0,1,2,3,4,5,..........  \nonumber \\
N_{n-\left( \mu +1\right) } &=&0,1,2,3,4,.......,N_{n-\mu }\text{ , \ \ with
\ }\mu =2,3,4,...,n-4\text{ \ ,}  \nonumber \\
\left( N_{2}=\right) l &=&0,1,2,3,4,.........N_{3\text{ }}\text{ \ and} 
\nonumber \\
m &=&-l,-l+1,.......,0,1,2,.......l\text{ \ .}
\end{eqnarray}

We now impose the Dirichlet boundary condition for our problem i.e.

\begin{equation}
\Phi _{\lambda }\left( x\right) \left| _{\chi =\chi _{0}}\right. =0\text{ \
at \ }\chi _{0}=\frac{\pi }{2}\text{ \ .}
\end{equation}

Using the wave function given Eq. (46) we immediately get integer values for 
$N$, which could be tabulated as in table II.. The new eigenvalues \ $\omega
_{k}=\frac{N\left( k\right) }{R_{0}}$ and the degenaracy $g_{k}$ could be
written as

\begin{eqnarray}
\omega _{k} &=&\frac{1}{R_{0}}\left( k+\frac{\left( n-2\right) }{2}\right) 
\text{ , \ }k=0,1,2,3,4,....  \nonumber \\
g_{k} &=&\left\{ 
\begin{array}{c}
\text{ }k=1,\text{ }N_{n-2}=0 \\ 
k=2,\text{ }N_{n-2}=1 \\ 
k=3,\text{ }N_{n-2}=0,2 \\ 
k=4,\text{ }N_{n-2}=1,3 \\ 
k=5,\text{ }N_{n-2}=0,2,4 \\ 
\vdots \text{ \ \ \ \ \ \ \ \ \ \ \ \ \ }\vdots \text{ }
\end{array}
\right\} \text{ }.%
\mathop{\displaystyle\sum}%
\limits_{N_{n-3}=0}^{N_{n-2}}\text{ }%
\mathop{\displaystyle\sum}%
_{N_{n-4}=0}^{N_{n-3}}\text{.....}%
\mathop{\displaystyle\sum}%
_{N_{4}=0}^{N_{5}}%
\mathop{\displaystyle\sum}%
_{N_{3}=0}^{N_{4}}%
\mathop{\displaystyle\sum}%
_{l=0}^{N_{3}}\left( 2l+1\right) \text{ \ .}
\end{eqnarray}

The new eigenfunctions Eq. (46) are still the Gegenbauer polnomials, but the
only ones that satisfy the boundary condition $\Phi _{\lambda }\left(
x\right) \left| _{\chi =\chi _{0}}\right. =0$ are picked: hence table II.
could be formed by just taking the second, fourth, etc. diagonal elements of
the table I..

Now we impose the Neumann boundary condition, which is given by

\begin{equation}
n^{\alpha }\text{ }\nabla _{\alpha }\Phi _{\lambda }\left( x\right) =0\text{
\ at }\chi _{0}=\frac{\pi }{2}\text{ ,}
\end{equation}

where $n^{\alpha }$ is the outward normal to the boundary. Hence for the
Neumann boundary condition

\begin{equation}
\frac{d}{d\theta }\Phi _{\lambda }\left( x\right) \mid _{\chi =\chi _{0}}=0%
\text{ \ at \ }\chi _{0}=\frac{\pi }{2}\text{ .}
\end{equation}

We obtain the $N$ values presented in table III. The new eigenvalues $\omega
_{k}$ and the degeneracy could be written as

\begin{eqnarray}
\omega _{k} &=&\frac{1}{R_{0}}\left( k+\frac{\left( n-2\right) }{2}\right) 
\text{ , \ }k=0,1,2,3,4,....  \nonumber \\
g_{k} &=&\left\{ 
\begin{array}{c}
\text{ }k=0,\text{ }N_{n-2}=0 \\ 
k=1,\text{ }N_{n-2}=1 \\ 
k=2,\text{ }N_{n-2}=0,2 \\ 
k=3,\text{ }N_{n-2}=1,3 \\ 
k=4,\text{ }N_{n-2}=0,2,4 \\ 
\vdots \text{ \ \ \ \ \ \ \ \ \ \ \ \ \ }\vdots \text{ }
\end{array}
\right\} \text{ }.%
\mathop{\displaystyle\sum}%
\limits_{N_{n-3}=0}^{N_{n-2}}\text{ }%
\mathop{\displaystyle\sum}%
_{N_{n-4}=0}^{N_{n-3}}\text{.....}%
\mathop{\displaystyle\sum}%
_{N_{4}=0}^{N_{5}}%
\mathop{\displaystyle\sum}%
_{N_{3}=0}^{N_{4}}%
\mathop{\displaystyle\sum}%
_{l=0}^{N_{3}}\left( 2l+1\right) \text{ \ .}
\end{eqnarray}

Notice that the $N$ values in table III. could be obtained from the full
universe case in table I. by taking the first, third, etc. diagonal
elements. Once again, the eigenfunctions are the Gegenbauer polnomials, but
only the ones that satisfy the Neumann boundary condition are allowed.

The Green's function that we aim to calculate is defined by

\begin{equation}
\left\langle 0\left| \Phi \left( x\right) \Phi \left( x^{\prime }\right)
\right| 0\right\rangle =%
\mathop{\displaystyle\sum}%
\limits_{\lambda }\text{ }\Phi _{\lambda }\left( x\right) \text{ }\Phi
_{\lambda }^{\ast }\left( x^{\prime }\right) \text{ ,}
\end{equation}

where $\Phi _{\lambda }\left( x\right) $ is the solution of the wave
equation Eq. (45) with the appropriate boundary conditions Eqs. (48) and
(51). Using the wave function given in Eq. (46), one obtains

\bigskip

\bigskip

\bigskip

\begin{eqnarray}
\left\langle 0\left| \Phi \left( x\right) \Phi \left( x^{\prime }\right)
\right| 0\right\rangle &=&\frac{2}{\pi R_{0}^{n-2}}%
\mathop{\displaystyle\sum}%
\limits_{\lambda }e^{-i\frac{\Delta t}{R_{0}}\left( k+\frac{\left(
n-2\right) }{2}\right) }\text{ }2^{2N_{n-2}+n-4}\left\{ \Gamma \left(
N_{n-2}+\frac{\left( n-2\right) }{2}\right) \right\} ^{2}  \nonumber \\
&&\frac{\left( k-N_{n-2}\right) !}{\Gamma \left( N_{n-2}+k+n-2\right) }\text{%
sin}^{N_{n-2}}\chi \text{sin}^{N_{n-2}}\chi ^{\prime }\text{ }%
C_{k-N_{n-2}}^{N_{n-2}+\frac{\left( n-2\right) }{2}}\left( \cos \chi \right)
\medskip \text{ }C_{k-N_{n-2}}^{N_{n-2}+\frac{\left( n-2\right) }{2}}(\cos
\chi ^{\prime })  \nonumber \\
&&\text{\ \ \ \ \ \ }%
\mathop{\displaystyle\prod}%
\limits_{\mu =1}^{n-4}\left\{ 
\mathop{\displaystyle\sum}%
\limits_{N_{n-\mu -2}=0}^{N_{n-\mu -1}}2^{2N_{n-\mu -2}+n-\mu -3}\left\{
\Gamma \left( N_{n-\mu -2}+\frac{\left( n-\mu -2\right) }{2}\right) \right\}
^{2}\text{ }\right.  \nonumber \\
&&\left[ \frac{\left( N_{n-\mu -1}-N_{n-\mu -2}\right) !\left( N_{n-\mu -1}+%
\frac{\left( n-\mu -2\right) }{2}\right) }{\pi \Gamma \left( N_{n-\mu
-1}+N_{n-\mu -2}+n-\mu -2\right) }\right]  \nonumber \\
&&\text{\ \ \ }\left. \text{sin}^{N_{n-\mu -2}}\theta _{\mu }\text{ sin}%
^{N_{n-\mu -2}}\theta _{\mu }^{\prime }\text{ }C_{N_{n-\mu -1}-N_{n-\mu
-2}}^{N_{n-\mu -2}+\frac{\left( n-\mu -2\right) }{2}}\left( \cos \theta
_{\mu }\right) \medskip \text{ }C_{N_{n-\mu -1}-N_{n-\mu -2}}^{N_{n-\mu -2}+%
\frac{\left( n-\mu -2\right) }{2}}\left( \cos \theta _{\mu }^{\prime
}\right) \right\}  \nonumber \\
&&\text{\ \ \ }%
\mathop{\displaystyle\sum}%
\limits_{m=-l}^{l}\text{ }Y_{lm\text{ }}\left( \theta _{n-3},\phi \right) 
\text{ }Y_{lm}^{\ast }\left( \theta _{n-3}^{\prime },\phi ^{\prime }\right) 
\text{ .}
\end{eqnarray}

$\lambda (N_{n-2},k)$ stands the new eigenvalues for the Dirichlet and
Neumann boundary conditions are presented in tables II. and III..

We define the Green's function for the Dirichlet and Neumann boundary
conditions as

\begin{eqnarray}
\left\langle 0\left| \Phi _{D}\left( x\right) \Phi _{D}\left( x^{\prime
}\right) \right| 0\right\rangle &=&D_{D}\left( x,x^{\prime }\right) \text{ ,
and\ }  \nonumber \\
\left\langle 0\left| \Phi _{N}\left( x\right) \Phi _{N}\left( x^{\prime
}\right) \right| 0\right\rangle &=&D_{N}\left( x,x^{\prime }\right) \text{ .}
\end{eqnarray}

We easily see that

\begin{equation}
D_{D}\left( x,x^{\prime }\right) +D_{N}\left( x,x^{\prime }\right) =2\text{ }%
D\text{ }\left( x,x^{\prime }\right) \text{ .}
\end{equation}

Where $D\left( x,x^{\prime }\right) =\left\langle 0\left| \Phi \left(
x\right) \Phi \left( x^{\prime }\right) \right| 0\right\rangle $ is the
Green's function for the n-dimensional universe which given in Eq. (40).
Since the sum of the Dirichlet and Neumann eigenvalues for the half universe
which is depicted in tables II. and III. is equal to full n-dimensional
universe eigenvalues given in table I.. The factor of $2$ comes from the
fact that the mode functions used in $D\left( x,x^{\prime }\right) $ are
normalized with respect to the n-dimensional universe.

\bigskip To evaluate $D_{D}\left( x,x^{\prime }\right) $ and $D_{N}\left(
x,x^{\prime }\right) $ explicitly we find it convenient to write them as

\begin{eqnarray}
D_{D}\left( x,x^{\prime }\right) &=&D\left( x,x^{\prime }\right) -\frac{1}{2}%
\left[ D_{N}\left( x,x^{\prime }\right) -D_{D}\left( x,x^{\prime }\right) %
\right] \text{ , and} \\
D_{N}\left( x,x^{\prime }\right) &=&D\left( x,x^{\prime }\right) +\frac{1}{2}%
\left[ D_{N}\left( x,x^{\prime }\right) -D_{D}\left( x,x^{\prime }\right) %
\right] \text{ .}
\end{eqnarray}

\bigskip The second term on the right hand side of Eqs. (57) and (58)\ could
be written as $D_{B}\left( x,x^{\prime }\right) =\frac{1}{2}\left[
D_{N}\left( x,x^{\prime }\right) -D_{D}\left( x,x^{\prime }\right) \right] $
. A closed expression for the $D\left( x,x^{\prime }\right) $ is already
known which given in Eq. (40) and could be written as

\begin{equation}
\left( G^{\left( +\right) }\left( x,x^{\prime }\right) \text{ =}\right) 
\text{ \ }D\left( x,x^{\prime }\right) \text{\ }=\frac{\alpha _{n}}{4\pi
^{n-2}R_{0}^{n-2}}\frac{1}{\left( \cos \frac{\Delta t}{R_{0}}-\cos \frac{%
\Delta s_{n-2}}{R_{0}}\right) ^{\frac{\left( n-2\right) }{2}}}\text{ \ .}
\end{equation}

\bigskip Where $\Delta t=t-t^{\prime }$ , and

\begin{equation}
\text{\ cos}\frac{\Delta s_{n-2}}{R_{0}}=\cos \chi \cos \chi ^{\prime }+\sin
\chi \sin \chi ^{\prime }\cos \gamma _{n-3}\text{ .}
\end{equation}

Here $\cos \gamma _{n-3}$ and $\alpha _{n}$ are given by Eq. (38) and Eq.
(39), respectively.

Using the information given in tables II. and III., definitions of $%
D_{N}\left( x,x^{\prime }\right) $ and $D_{D}\left( x,x^{\prime }\right) $ ,
and addition theorem for the Gegenbauer polynomials one could write as

\begin{eqnarray}
\text{ \ }D_{B}\left( x,x^{\prime }\right) \text{\ } &=&\frac{\alpha _{n}}{%
4\pi ^{n-2}R_{0}^{n-2}}\text{ }\frac{\Gamma \left( n-3\right) }{\left\{
\Gamma \left( \frac{n-2}{2}\right) \right\} ^{2}}\left[ e^{-i\frac{\left(
n-2\right) \Delta t}{2R_{0}}}\frac{\left\{ \Gamma \left( \frac{n-2}{2}%
\right) \right\} ^{2}}{\Gamma \left( n\text{ }-2\right) }\text{ }\left(
n-3\right) \right.   \nonumber \\
&&C_{0}^{\frac{\left( n-2\right) }{2}}\left( \cos \chi \right) \text{ }%
C_{0}^{\frac{\left( n-2\right) }{2}}\left( \cos \chi ^{\prime }\right) \text{
}C_{0}^{\frac{\left( n-3\right) }{2}}\left( \cos \gamma _{n-3}\right)  
\nonumber \\
&&+e^{-i\frac{n\Delta t}{2R_{0}}\text{ }}\text{ }\left\{ \frac{2\left\{
\Gamma \left( \frac{n}{2}\right) \right\} ^{2}}{\Gamma \left( n\text{ }%
\right) }\left( n-1\right) \text{ }\sin \chi \sin \chi ^{\prime }\right.  
\nonumber \\
&&C_{0}^{\frac{n}{2}}\left( \cos \chi \right) \text{ }C_{0}^{\frac{n}{2}%
}\left( \cos \chi ^{\prime }\right) \text{ }C_{1}^{\frac{\left( n-3\right) }{%
2}}\left( \cos \gamma _{n-3}\right)   \nonumber \\
&&\left. -\frac{\left\{ \Gamma \left( \frac{n-2}{2}\right) \right\} ^{2}}{%
\Gamma \left( n\text{ -1}\right) }\left( n-3\right) \text{ }C_{1}^{\frac{%
\left( n-2\right) }{2}}\left( \cos \chi \right) \text{ }C_{1}^{\frac{\left(
n-2\right) }{2}}\left( \cos \chi ^{\prime }\right) \text{ }C_{0}^{\frac{%
\left( n-3\right) }{2}}\left( \cos \gamma _{n-3}\right) \right\}   \nonumber
\\
&&+e^{-i\frac{\left( n+2\right) \Delta t}{2R_{0}}\text{ }}\text{ }\left\{ 
\frac{2!\left\{ \Gamma \left( \frac{n-2}{2}\right) \right\} ^{2}}{\Gamma
\left( n\text{ }\right) }\left( n-3\right) \text{ }C_{2}^{\frac{\left(
n-2\right) }{2}}\left( \cos \chi \right) \right.   \nonumber \\
&&C_{2}^{\frac{\left( n-2\right) }{2}}\left( \cos \chi ^{\prime }\right) 
\text{ }C_{0}^{\frac{\left( n-3\right) }{2}}\left( \cos \gamma _{n-3}\right) 
\nonumber \\
&&-\frac{2^{2}\left\{ \Gamma \left( \frac{n}{2}\right) \right\} ^{2}1!}{%
\Gamma \left( n+1\text{ }\right) }\left( n-1\right) \text{ }\sin \chi \sin
\chi ^{\prime }  \nonumber \\
&&C_{1}^{\frac{n}{2}}\left( \cos \chi \right) \text{ }C_{1}^{\frac{n}{2}%
}\left( \cos \chi ^{\prime }\right) \text{ }C_{1}^{\frac{\left( n-3\right) }{%
2}}\left( \cos \gamma _{n-3}\right)   \nonumber \\
&&+\frac{2^{4}\left\{ \Gamma \left( \frac{n+6}{2}\right) \right\} ^{2}0!}{%
\Gamma \left( n+2\text{ }\right) }\left( n+1\right) \sin ^{2}\chi \sin
^{2}\chi ^{\prime }C_{0}^{\frac{\left( n+2\right) }{2}}\left( \cos \chi
\right)   \nonumber \\
&&\left. C_{0}^{\frac{\left( n+2\right) }{2}}\left( \cos \chi ^{\prime
}\right) \text{ }C_{2}^{\frac{\left( n-3\right) }{2}}\left( \cos \gamma
_{n-3}\right) \right\} +.....................\left. {}\right] \text{ \ .}
\end{eqnarray}

\bigskip Using the symmetry properties of Gegenbauer polynomials

\begin{equation}
C_{k-N_{n-2}}^{N_{n-2}+\frac{\left( n-2\right) }{2}}\left( -\cos \chi
^{\prime }\right) =\left( -1\right) ^{k-N_{n-2}}\text{ }%
C_{k-N_{n-2}}^{N_{n-2}+\frac{\left( n-2\right) }{2}}\left( \cos \chi
^{\prime }\right) \text{ ,}
\end{equation}

we could now rewrite Eq. (61)\ \ as

\begin{eqnarray}
\text{\ }D_{B}\left( x,x^{\prime }\right) \text{\ } &=&\frac{\alpha _{n}}{%
4\pi ^{n-2}R_{0}^{n-2}}\text{ }\frac{\Gamma \left( n-3\right) }{\left\{
\Gamma \left( \frac{n-2}{2}\right) \right\} ^{2}}\text{ }%
\mathop{\displaystyle\sum}%
\limits_{k=0}^{\infty }\text{ e}^{-i\omega _{k}\Delta t}\text{ }%
\mathop{\displaystyle\sum}%
\limits_{N_{n-2}=0}^{k}\text{ }2^{N_{n-2}}\frac{\left\{ \Gamma \left(
N_{n-2}+\frac{\left( n-2\right) }{2}\right) \right\} ^{2}}{\Gamma \left(
N_{n-2}+k+\left( n-2\right) \right) }  \nonumber \\
&&\left( 2N_{n-2}+n-3\right) \text{ }\sin ^{N_{n-2}}\chi \sin ^{N_{n-2}}\chi
^{\prime }  \nonumber \\
&&C_{k-N_{n-2}}^{N_{n-2}+\frac{\left( n-2\right) }{2}}\left( \cos \chi
\right) \text{ }C_{k-N_{n-2}}^{N_{n-2}+\frac{\left( n-2\right) }{2}}\left(
-\cos \chi ^{%
{\acute{}}%
}\right) \text{ }C_{N_{n-2}}^{\frac{\left( n-3\right) }{2}}\left( \cos
\gamma _{n-3}\right) \text{ \ .}
\end{eqnarray}

Using the addition theorem for the Gegenbauer polynomials, one obtains

\begin{equation}
D_{B}\left( x,x^{\prime }\right) \text{\ }=\frac{\alpha _{n}}{4\pi
^{n-2}R_{0}^{n-2}}\text{ }%
\mathop{\displaystyle\sum}%
\limits_{k=0}^{\infty }\text{ e}^{-i\omega _{k}\Delta t}\text{ C}_{k}^{\frac{%
\left( n-2\right) }{2}}\left( \cos \overline{\gamma _{n-2}}\right) \text{ ,}
\end{equation}

where

\begin{equation}
\cos \overline{\gamma _{n-2}}=-\cos \chi \cos \chi ^{\prime }+\sin \chi \sin
\chi ^{\prime }\cos \gamma _{n-3}\text{ .}
\end{equation}

Letting $\Delta t\rightarrow \Delta t-i\epsilon $ where $\epsilon
\rightarrow 0$ , then $\left| e^{-i\frac{\Delta t}{R_{0}}}\right| <1$ . Now
the sum in Eq. (64) could be evaluated using the generating function of
Gegenbauer polynomials, one obtains

\begin{equation}
D_{B}\left( x,x^{\prime }\right) \text{\ }=\frac{\alpha _{n}}{4\pi
^{n-2}R_{0}^{n-2}}\frac{1}{\left( \cos \frac{\Delta t}{R_{0}}-\cos \frac{%
\Delta \widetilde{s_{n-2}}}{R_{0}}\right) ^{\frac{\left( n-2\right) }{2}}}%
\text{ ,}
\end{equation}

where

\bigskip 
\begin{eqnarray}
\left( \cos \widetilde{\gamma _{n-2}}=\right) \text{ \ cos}\frac{\Delta 
\widetilde{s_{n-2}}}{R_{0}} &=&-\cos \chi \cos \chi ^{\prime }+\sin \chi
\sin \chi ^{\prime }\cos \gamma _{n-3}\text{ ,}  \nonumber \\
\cos \text{ }\gamma _{n-\mu -2} &=&\cos \theta _{\mu }\cos \theta _{\mu
}^{\prime }+\sin \theta _{\mu }\sin \theta _{\mu }^{\prime }\cos \gamma
_{n-\mu -3}  \nonumber \\
\text{with }\mu &=&2,3,4,....,n-4\text{ , and}  \nonumber \\
\cos \text{ }\gamma _{1} &=&\cos \theta _{n-3}\cos \theta _{n-3}^{\prime
}+\sin \theta _{n-3}\sin \theta _{n-3}^{\prime }\cos \left( \phi -\phi
^{\prime }\right) \text{ .}
\end{eqnarray}

We could now write the complete Green's functions $D_{N}\left( x,x^{\prime
}\right) $ and $D_{D}\left( x,x^{\prime }\right) $ as

\begin{eqnarray}
D_{N}\left( x,x^{\prime }\right) \text{\ } &=&\frac{\alpha _{n}}{4\pi
^{n-2}R_{0}^{n-2}}\left[ \frac{1}{\left( \cos \frac{\Delta t}{R_{0}}-\cos 
\frac{\Delta s_{n-2}}{R_{0}}\right) ^{\frac{\left( n-2\right) }{2}}}+\frac{1%
}{\left( \cos \frac{\Delta t}{R_{0}}-\cos \frac{\Delta \widetilde{s_{n-2}}}{%
R_{0}}\right) ^{\frac{\left( n-2\right) }{2}}}\right] \text{ ,} \\
D_{D}\left( x,x^{\prime }\right) \text{\ } &=&\frac{\alpha _{n}}{4\pi
^{n-2}R_{0}^{n-2}}\left[ \frac{1}{\left( \cos \frac{\Delta t}{R_{0}}-\cos 
\frac{\Delta s_{n-2}}{R_{0}}\right) ^{\frac{\left( n-2\right) }{2}}}-\frac{1%
}{\left( \cos \frac{\Delta t}{R_{0}}-\cos \frac{\Delta \widetilde{s_{n-2}}}{%
R_{0}}\right) ^{\frac{\left( n-2\right) }{2}}}\right] \text{ .}
\end{eqnarray}

Where $\cos \frac{\Delta s_{n-2}}{R_{0}}$ and $\cos \frac{\Delta \widetilde{%
s_{n-2}}}{R_{0}}$ are given by Eq. (39) and Eq. (67), respectively.

We also note that these Green's functions are naturally identical to the
Green's functions that one could get by using the image method, which always
works on the double manifold defined by \cite{kennedy,dow1,m2}

\begin{equation}
M\cup \partial M\cup M^{\ast }\text{ .}
\end{equation}

$M$ is the physical space inside the boundary, $\partial M$ is the boundary
and $M^{\ast }$ is the dual space obtained by reflecting the physical space
about the boundary. The field is confined to the region defined by $M$ and
satisfies the boundary conditions Eqs. (48) and (50). To satisfy the
boundary condition on $\partial M$ it is sufficient to locate an image
charge in the unphysical dual region $M^{\ast }$. The Green's function on $%
M\cup \partial M$ is then given by

\begin{equation}
G\left( x,x^{\prime }\right) =D\left( x,x^{\prime }\right) \pm D\left( x,%
\widetilde{x^{\prime }}\right) \text{ ,}
\end{equation}

where $-\left( +\right) $ refers to the Dirichlet (Neumann) boundary
condition, $D\left( x,x^{\prime }\right) $ is the Green's function for the
double manifold and $\widetilde{x^{\prime }}$ is the image of $x^{\prime }.$
The key element in the success of image method is that the solutions of Eq.
(45) found in the double manifold will have distinct parity (even or odd)
under transformation $x\rightarrow \widetilde{x}$. Notice that in Eq. (68)
and Eq. (69) the first term is the Green's function for the double manifold,
which in this case is the full $S^{n-1},$ and the $\left( \cos \widetilde{%
\gamma _{n-2}}=\right) $ \ cos$\frac{\Delta \widetilde{s_{n-2}}}{R_{0}}$ \
in the second term could be rewritten as

\begin{equation}
\left( \cos \widetilde{\gamma _{n-2}}=\right) \text{ \ cos}\frac{\Delta 
\widetilde{s_{n-2}}}{R_{0}}=-\cos \chi \cos \chi ^{\prime }+\sin \chi \sin
\chi ^{\prime }\cos \gamma _{n-3}\text{ ,}
\end{equation}

which could be obtained from cos$\frac{\Delta s_{n-2}}{R_{0}}$ by
substituting the image of $x^{\prime }$ i.e. by replacing $\chi ^{\prime
}\rightarrow \pi -\chi ^{\prime }.$

\section{SUMMARY}

We constructed the Green's function by using the eigenfunctions, which are
obtained by solving the wave equation for a conformal scalar field in a
n-dimensional closed, static universe and with the appropriate boundary
conditions (Dirichlet and Neumann). Even though we use the global topology
of \ the n-dimensional universe to construct the Green's function, our
result is locally true. Our result is interesting, to understand the
correspondence between the sign of the Casimir energy and different manifold
structure in closed topology of the universe.

\bigskip

Our result is an agreement with the results obtained by the method of
images. TheGreen's function is calculated for an isolated sphere, where the
interior geometry is given by Eq. (44) (half space universe). The global
topology of the universe as well as the outside geometry are not important
and have no effect on our result. This very important for applications of
the Casimir effect to the bag models. Currently we are investigating the
Casimir effect for a massless conformal scalar field in a n-dimensional
closed, static universe and a half space with Dirichlet and Neumann boundary
conditions.

\bigskip

\bigskip

\bigskip

\bigskip

\bigskip

\bigskip

\bigskip

\bigskip

\bigskip

\bigskip

\bigskip

Table I. : $N$ $\left( k,N_{n-2}\right) $ values for the full space case

\bigskip

\bigskip

\bigskip

. 
\begin{tabular}[t]{|l|l|l|l|l|l|l|l|}
\hline
$N_{n-2}\diagdown k$ & $0$ & $1$ & $2$ & $3$ & 4 & $5$ & $\cdots $ \\ \hline
$0$ & $\frac{n-2}{2}$ & $\frac{n}{2}$ & $\frac{n+2}{2}$ & $\frac{n+4}{2}$ & $%
\frac{n+6}{2}$ & $\frac{n+8}{2}$ & $\cdots $ \\ \hline
$1$ &  & $\frac{n}{2}$ & $\frac{n+2}{2}$ & $\frac{n+4}{2}$ & $\frac{n+6}{2}$
& $\frac{n+8}{2}$ & $\cdots $ \\ \hline
$2$ &  &  & $\frac{n+2}{2}$ & $\frac{n+4}{2}$ & $\frac{n+6}{2}$ & $\frac{n+8%
}{2}$ & $\cdots $ \\ \hline
$3$ &  &  &  & $\frac{n+4}{2}$ & $\frac{n+6}{2}$ & $\frac{n+8}{2}$ & $\cdots 
$ \\ \hline
$4$ &  &  &  &  & $\frac{n+6}{2}$ & $\frac{n+8}{2}$ & $\cdots $ \\ \hline
$5$ &  &  &  &  &  & $\frac{n+8}{2}$ & $\cdots $ \\ \hline
$\vdots $ &  &  &  &  &  &  & $\ddots $ \\ \hline
\end{tabular}

\bigskip

\bigskip

\bigskip

Table II.: $N\left( k,N_{n-2}\right) $ values for the Dirichlet boundary case

\bigskip

. . 
\begin{tabular}[t]{|l|l|l|l|l|l|l|l|}
\hline
$N_{n-2}\diagdown k$ & $0$ & $1$ & $2$ & $3$ & 4 & $5$ & $\cdots $ \\ \hline
$0$ &  & $\frac{n}{2}$ &  & $\frac{n+4}{2}$ &  & $\frac{n+8}{2}$ &  \\ \hline
$1$ &  &  & $\frac{n+2}{2}$ &  & $\frac{n+6}{2}$ &  & $\ddots $ \\ \hline
$2$ &  &  &  & $\frac{n+4}{2}$ &  & $\frac{n+8}{2}$ &  \\ \hline
$3$ &  &  &  &  & $\frac{n+6}{2}$ &  & $\ddots $ \\ \hline
$4$ &  &  &  &  &  & $\frac{n+8}{2}$ &  \\ \hline
$5$ &  &  &  &  &  &  & $\ddots $ \\ \hline
$\vdots $ &  &  &  &  &  &  &  \\ \hline
\end{tabular}

\bigskip

\bigskip

\bigskip

\bigskip

\bigskip

\bigskip

\bigskip

\bigskip

\bigskip

\bigskip

\bigskip

\bigskip

\bigskip

\bigskip

\bigskip

\bigskip

\bigskip

\bigskip

\bigskip

\bigskip

\bigskip

Table III.: $N\left( k,N_{n-2}\right) $ values for the Neumann boundary case.

\bigskip

\bigskip

\bigskip

\bigskip

\bigskip

\bigskip

\bigskip

\bigskip

\bigskip

\bigskip

\begin{tabular}[t]{|l|l|l|l|l|l|l|l|}
\hline
$N_{n-2}\diagdown k$ & $0$ & $1$ & $2$ & $3$ & 4 & $5$ & $\cdots $ \\ \hline
$0$ & $\frac{n-2}{2}$ &  & $\frac{n+2}{2}$ &  & $\frac{n+6}{2}$ &  & $\ddots 
$ \\ \hline
$1$ &  & $\frac{n}{2}$ &  & $\frac{n+4}{2}$ &  & $\frac{n+8}{2}$ &  \\ \hline
$2$ &  &  & $\frac{n+2}{2}$ &  & $\frac{n+6}{2}$ &  & $\ddots $ \\ \hline
$3$ &  &  &  & $\frac{n+4}{2}$ &  & $\frac{n+8}{2}$ &  \\ \hline
$4$ &  &  &  &  & $\frac{n+6}{2}$ &  & $\ddots $ \\ \hline
$5$ &  &  &  &  &  & $\frac{n+8}{2}$ &  \\ \hline
$\vdots $ &  &  &  &  &  &  & $\ddots $ \\ \hline
\end{tabular}

\end{document}